\documentclass[pdflatex,sn-mathphys-num]{sn-jnl}

\usepackage{graphicx}%
\usepackage{multirow}%
\usepackage{amsmath,amssymb,amsfonts}%
\usepackage{amsthm}%
\usepackage{mathrsfs}%
\usepackage[title]{appendix}%
\usepackage{textcomp}%
\usepackage{manyfoot}%
\usepackage{booktabs}%
\usepackage{algorithm}%
\usepackage{algorithmicx}%
\usepackage{algpseudocode}%
\usepackage{listings}%

\usepackage{url}
\usepackage[table,dvipsnames]{xcolor}
\definecolor{left}{HTML}{172869}     
\definecolor{left-center}{HTML}{0076BB}           
\definecolor{center}{HTML}{399043}         
\definecolor{right-center}{HTML}{AF6125}    
\definecolor{right}{HTML}{881C00}          
\definecolor{extreme-right}{HTML}{762A83} 

\usepackage{multirow}
\usepackage{adjustbox}
\usepackage{makecell}
\usepackage{booktabs}
\usepackage{longtable}

\usepackage{caption}
\usepackage{stfloats}
\usepackage{amsmath}

\usepackage{graphicx}
\usepackage{hyperref}

\begin{document}

\title[Article Title]{Ideological Fragmentation of the Social Media Ecosystem: From echo chambers to echo platforms}

\author*[1]{\fnm{Edoardo} \sur{Di Martino}}\email{edoardo.dimartino@uniroma1.it}
\author[2]{\fnm{Alessandro} \sur{Galeazzi}}\email{alessandro.galeazzi@unipd.it}
\author[3, 4]{\fnm{Michele} \sur{Starnini}}\email{michele.starnini@upf.edu}
\author[5]{\fnm{Walter} \sur{Quattrociocchi}}\email{walter.quattrociocchi@uniroma1.it}
\author[5]{\fnm{Matteo} \sur{Cinelli}}\email{matteo.cinelli@uniroma1.it}

\affil*[1]{\orgdiv{Department of Social Sciences and Economics}, \orgname{Sapienza University of Rome}, \orgaddress{\street{P.le Aldo Moro, 5}, \postcode{00185}, \state{Rome}, \country{Italy}}}

\affil[2]{\orgdiv{Department of Mathematics}, \orgname{University of Padova}, \orgaddress{\street{Via Trieste, 63}, \postcode{ 35121 }, \state{Padova}, \country{Italy}}}

\affil[3]{\orgdiv{Department of Engineering}, \orgname{Universitat Pompeu Fabra}, \orgaddress{\postcode{08018}, \state{Barcelona}, \country{Spain}}}
\affil[4]{\orgname{CENTAI},\postcode{10138}, \state{Torino}, \country{Italy}}

\affil[5]{\orgdiv{Department of Computer Science}, \orgname{Sapienza University of Rome}, \orgaddress{\street{ Viale Regina Elena, 295}, \postcode{00161}, \state{Rome}, \country{Italy}}}

\abstract{The entertainment-driven nature of social media encourages users to engage with like-minded individuals and consume content aligned with their beliefs, limiting exposure to diverse perspectives. Simultaneously, users migrate between platforms, either due to moderation policies like de-platforming or in search of environments better suited to their preferences. These dynamics drive the specialization of the social media ecosystem, shifting from internal echo chambers to ``echo platforms''—entire platforms functioning as ideologically homogeneous niches. 
To systematically analyze this phenomenon in political discussions, we propose a quantitative approach based on three key dimensions: platform centrality, news consumption, and user base composition. 
We analyze 117 million posts related to the 2020 US Presidential elections from nine social media platforms—Facebook, Reddit, Twitter, YouTube, BitChute, Gab, Parler, Scored, and Voat. 
Our findings reveal significant differences among platforms in their centrality within the ecosystem, the reliability of circulated news, and the ideological diversity of their users, highlighting a clear divide between mainstream and alt-tech platforms. 
The latter occupy a peripheral role, feature a higher prevalence of unreliable content, and exhibit greater ideological uniformity. These results highlight the key dimensions shaping the fragmentation and polarization of the social media landscape.}

\maketitle

\section{Introduction}

Social media have become a major source of news and opinions for many, reshaping how people access information and engage in public discussions. 
This shift coincides with a decline in trust in traditional media~\cite{gallup2018indicators, nic2018reuters} and is driven by platforms that prioritize entertainment and user engagement over informational accuracy~\cite{voorveld2018engagement, sangiorgio2024followers, cinelli2021echo, etta2023characterizing}. In this context, the combination of recommendation algorithms, designed to maximize engagement, and user preferences may encourage individuals to join tightly-knit groups of like-minded peers and consume mostly content that aligns with their pre-existing beliefs, reinforcing shared perspectives and potentially limiting exposure to diverse viewpoints~\cite{bessi2015science, cinelli2021echo, terren2021echo}. This behavior has been closely linked to heightened exposure to hate speech~\cite{avalle2024persistent, shandwick26krc, league2020online, tahmasbi2021go}, political polarization~\cite{bail2018exposure, kubin2021role, falkenberg2022growing}, and the formation of ideologically homogeneous clusters of users, called echo chambers~\cite{cinelli2021echo, cinus2022effect, del2016spreading, diaz2023disinformation, bovet2022organization}. Over time, the echo chamber effect may drive groups towards more extreme positions~\cite{sunstein1999law, del2016echo} and influence user behavior across platforms~\cite{cinelli2021echo, hobolt2024polarizing, monti2023online} going even beyond digital interactions~\cite{mccoy2019polarization, mccoy2022happens}.

To address these challenges, platforms have increasingly implemented moderation policies aimed at managing tight-knit, ``problematic'' communities~\cite{grimmelmann2015virtues}. However, moderation policies, such as de-platforming, often push users toward minimally regulated platforms~\cite{cinus2022effect, ali2021understanding, monti2023online, mekacher2023systemic, cima2024great}. At the same time, users may spontaneously leave platforms where they find limited alignment with their preferences or feel overly exposed to toxic content, seeking alternatives. For example, many users migrated from Twitter to Mastodon~\cite{cava2023drivers} or BlueSky~\cite{quelle2025bluesky, failla2024m} following changes in platform management.
When users leave mainstream platforms, they often migrate to less popular digital spaces, commonly referred to as ``alt-tech'' or ``fringe'' platforms~\cite{newell2016user, horta2021platform, monti2023online}. Examples include Gab, Parler, BitChute, and Rumble, which have emerged as counterparts to mainstream services like Twitter, Reddit, YouTube, and Facebook. These platforms frequently market themselves as champions of free speech, host alt-right or extremist content, and appeal to users who perceive mainstream platforms as ideologically biased or overly regulated. While their user bases are relatively small compared to mainstream platforms, these alt-tech environments can play a non-negligible role in shaping societal events, serving as important hubs for specific ideological groups, facilitating radicalization, amplifying fringe narratives, contributing to online fragmentation, and creating pathways from online discourse to offline action. For example, Parler was used to coordinate the Capitol Hill assault during the 2020 U.S. elections~\cite{GaisCruz2021, ABCNews2021}; similarly, Gab was linked to the Pittsburgh synagogue shooter~\cite{levenson2018gab}, and Voat became a refuge for QAnon communities banned from Reddit~\cite{monti2023online}.

Both spontaneous and forced migrations foster differentiation and specialization within the social media ecosystem, with platforms increasingly organized around specific user communities and shared content preferences~\cite{schulze2022far, cinelli2022conspiracy, horta2023deplatforming, de2023fringe}. These shifts represent a departure from the original vision of the world wide web as a globally interconnected network where information flows freely across borders~\cite{levy2005collective, howcroft1998utopia}. Instead, entire platforms are now defined by ideologically homogeneous communities and aligned content, creating niches where opposing perspectives are rarely encountered~\cite{cinelli2022conspiracy}.
This scenario—where an entire platform is characterized by a like-minded user base with limited exposure to diverse opinions—lays the foundation for the concept of an ``echo platform'', an extension of the echo chamber phenomenon to the level of entire digital ecosystems.

In this paper, we introduce three operational steps to characterize the fragmentation of the social media ecosystem, providing a methodological basis for identifying the role of different platforms. Using a dataset comprising over 117 million URLs collected from nine platforms—four mainstream (Facebook, Twitter, Reddit, YouTube) and five alt-tech fringe platforms (BitChute, Gab, Parler, Scored, Voat)—and nearly six million unique users, during the 2020 US presidential election, we define three axes to characterize platform roles: (i) centrality (central vs. peripheral), (ii) news consumption (reliable vs. questionable content), and (iii) user base composition (uniform vs. diverse).
In this context, we adopt the terms mainstream and alt-tech following the terminology commonly used in prior literature~\cite{ali2021understanding, horta2021platform}: we define alt-tech platforms as those that have been reported to systematically host alt-right or extremist content.
However, our goal is to move beyond these binary labels, providing a comprehensive data-driven characterization of the social media ecosystem.

Our analysis reveals a distinct pattern of fragmentation characterized by limited interactions. Alt-tech platforms exhibit significantly higher levels of ideological homogeneity, with user communities sharing content almost exclusively aligned with their dominant narratives. 
Interestingly, user behavior on Reddit more closely resembles that of alt-tech platforms rather than mainstream ones, featuring a uniformly composed user base that is, however, left-leaning, and playing a relatively peripheral role in cross-platform linking.
Furthermore, alt-tech platforms disproportionately amplify questionable content while showing a notable absence of reliable news sources compared to their mainstream counterparts.

\section*{Results}

We characterize the social media ecosystem along three key dimensions, providing detailed insights into the structural and functional distinctions between platforms.

\begin{itemize}
\item{\textbf{Centrality: Central vs Peripheral Role.}}
Platforms can be classified as either central or peripheral based on their position within the broader information ecosystem. 
Some platforms could serve as central nodes in the network connecting different platforms, due to their high connectivity.
Others may be detached from the core, operating as niches where users are less likely to point to other platforms' content.

\item{\textbf{News Consumption: Reliable vs. Questionable Sources.}}
The reliability of news circulating on platforms may vary significantly. 
Some platforms primarily feature content from reliable, well-established news sources that align with journalistic standards. 
Others display a higher prevalence of questionable content, including conspiracy theories and unverified narratives.

\item{\textbf{Political Leaning of Users: Uniform vs Diverse.}}
User base composition can differ substantially across platforms. 
Some platforms host a heterogeneous (broad) user base, encompassing individuals with diverse political stances and interests. 
Others are characterized by a homogeneous (narrow) user base where like-minded individuals share similar ideas and narratives. 
\end{itemize}

The systematic characterization of these three dimensions provides a framework for understanding the role of different platforms in the social media ecosystem. Our analysis is based on a comprehensive dataset of URLs collected from nine social media platforms. 

\begin{table*}[b]
    \centering
    \renewcommand{\arraystretch}{1.5} 
     \setlength{\tabcolsep}{0.9\tabcolsep}
    \caption{Dataset Statistics and Evaluation Metrics by Platform. For each platform, we report the following metrics: number of unique users ($N$), number of URL links ($n_u$), number of unique domains linked ($n_d$), PageRank centrality ($PR$), fraction of questionable sources shared ($q$), and the variance of the users' political leaning distribution ($\sigma^2$). The figures for the number of unique users, URLs, and unique domains are provided after the data preprocessing procedures.
    }
   \begin{tabular}{|l|c|c|c|c|c|c|c|c|}
        \hline
        \textbf{Platform} & $N$ & $n_u$ & $n_d$ & 
        $PR$ & 
        $q$ & 
        $\sigma^2$\\
        \hline
        Facebook   & 300k & 10.7M & 74k & 0.07 & 0.15 & 0.16 \\
        \hline
        Reddit     & 59k & 320k & 9k & 0.10 & 0.03 & 0.04 \\
        \hline
        Twitter    & 5M & 103M & 114k & 0.14 & 0.16 & 0.16 \\
        \hline
        YouTube    & 12k & 512k & 12k & 0.06 & 0.14 & 0.19 \\
        \hline
        BitChute   & 2k & 107k & 4k & 0.18 & 0.22 & 0.17 \\
        \hline
        Gab        & 15k & 453k & 7k & 0.16 & 0.65 & 0.06 \\
        \hline
        Parler     & 73k & 1.2M & 13k & 0.10 & 0.63 & 0.06 \\
        \hline
        Scored     & 41k & 868k & 11k & 0.04 & 0.78 & 0.04 \\
        \hline
        Voat       & 20k & 135k & 6k & 0.13 & 0.42 & 0.10 \\
        \hline
    \end{tabular}
    \label{tab:datasetbreakdown}
\end{table*}

As shown in Table \ref{tab:datasetbreakdown}, we include four ``mainstream" platforms (Facebook, Twitter, Reddit, and YouTube) and five ``alt-tech" platforms (BitChute, Gab, Parler, Scored, and Voat). 
For each platform, we collect only politically relevant posts, especially those related to the 2020 U.S. presidential campaign. 
Such content was obtained by querying for keywords related to the presidential candidates contained in the posts (Facebook, Twitter, Gab, and Parler) or the video titles or descriptions (YouTube and BitChute).
On Reddit, Voat, and Scored posts are often made of only URLs linking to the relevant piece of news, thus a keyword search would not be applicable. 
As such, we only selected content coming from political communities active during the 2020 U.S. presidential campaign, after a manual inspection of the posts. The time window for the analysis spans year 2020, with different months covered for each platform depending on data availability as detailed in the Materials and Methods section. While we report results based on the full data sets, we provide the analysis using uniformly restricted time frames in Figure S5-S9 of the Supplementary Information (SI), showing that the results remain consistent.

\subsection*{\textbf{Platform Centrality}}

To analyze the roles of different platforms within the social media ecosystem, we model it as a weighted, directed graph. 
In this network, nodes represent platforms, and an edge from platform $i$ to platform $j$ exists if an account of platform $i$ posted a URL linking directly to platform $j$.
The edge weight represents the number of URL links pointing from one platform to another. 
Therefore, the weighted adjacency matrix of this network (see SI, Fig. S1) can provide insights into how platforms redirect users and share content within the ecosystem. 
To account for variations in the size of the datasets collected from each platform (and their popularity), we divide the observed weights by the weights expected in a random network which preserves the total number of outgoing and incoming URL links for each platform (see section Materials and Methods for details). 

\begin{figure}[tbp]
  \centering
  \includegraphics[width=\columnwidth]{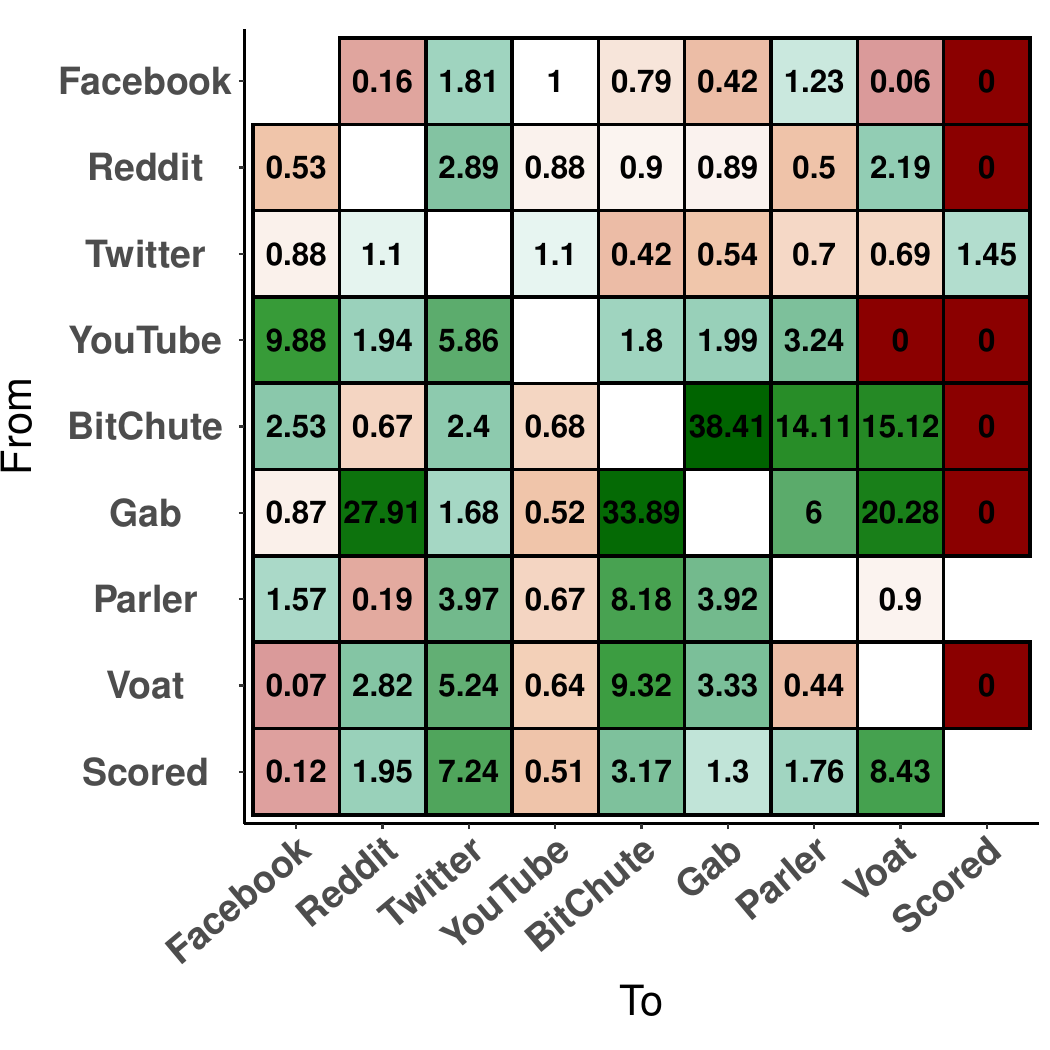}
  \caption{\textbf{Centrality.}
  Rescaled adjacency matrix showing the ratio between the observed and expected number of URLs pointing from one platform to another. 
  Green (red) cells indicate values greater (smaller) than one. 
  A value of 0 indicate no observed URLs. 
  }
  \label{fig:null_model}
\end{figure}

Fig.~\ref{fig:null_model} displays the ratio of observed edge weights to those expected under the null model. 
Ratios greater than one indicate that a platform links to another more frequently than expected in a random network, while ratios less than one indicate less frequent linking.

Fig.~\ref{fig:null_model} reveals distinct patterns of inter-platform linking: Twitter consistently receives more URLs than expected from all other platforms (third column), positioning it as a central node in the social media ecosystem. Surprisingly, Facebook receives fewer links than expected from other mainstream platforms (first column), except YouTube, likely due to the prevalence of self-promotional links in video descriptions. 
Alt-tech platforms generally receive fewer links than expected by mainstream ones, and are instead more tightly connected among themselves. 
Despite this observation, the third highest weight with respect to the null model is observed in the edge from Gab to Reddit, suggesting a potential content-sharing dynamic or migration pathway between these two platforms. 
In June 2020, indeed, Reddit banned the subreddit r/The\_Donald, a forum dedicated to President Trump's supporters with nearly $800,000$ users, due to repeated violations of the platform's rules against harassment, hate speech, and content manipulation~\cite{NPR2020}. 
Given the substantial presence of Trump supporters on Gab, this could suggest that links from Gab to Reddit were possibly posted as a means of mockery or demotion.

BitChute is linked much more than expected by all fringe platforms and slightly less by mainstream ones (fifth column), while the opposite is true for YouTube (fourth column). Furthermore, we observe a strong relationship between BitChute and Gab, as indicated by the two edges having the highest weights in the network. 
Finally, Scored does not receive incoming links from six of the eight other platforms, likely due to its limited popularity and low exposure, underscoring its peripheral role in the broader ecosystem. 

To better quantify the relative importance of each platform within the ecosystem, we computed their PageRank centrality from the rescaled adjacency matrix shown in Fig.~\ref{fig:null_model}. 
The results presented in Table~\ref{tab:datasetbreakdown} reveal that BitChute, Gab, and Twitter exhibit the highest scores, reflecting their prominent roles in the network. 
Voat, Parler, and Reddit display lower centrality scores but still maintain relatively strong connection patterns. Facebook, YouTube, and Scored, on the other hand, rank as the least central platforms in the ecosystem.
Therefore, after accounting for popularity, we found that not all mainstream platforms are central, while fringe platforms like Gab and BitChute are not necessarily peripheral.

\subsection*{\textbf{News Consumption}}

Next, we investigate whether the type of news and content shared differs significantly across the nine platforms by analyzing the news domains' political bias. 
To this aim, we use Media Bias/Fact Check (commonly referred to as MBFC \url{https://mediabiasfactcheck.com}), a widely referenced news rating agency that provides political bias and reliability labels for news outlets and other information producers \cite{stefanov2020predicting, cinelli2021echo, flamino2023political}. 
We extract the domains from the URLs shared across the nine platforms and match them with the political bias and reliability information obtained from MBFC. 
Domains not classified by MBFC are labeled as ``unreported'', though this designation does not necessarily imply an absence of political bias. Following previous research~\cite{flamino2023political}, domains categorized as ``extreme-left'' are excluded from the analysis due to their negligible presence in the dataset (see Materials and Methods for details regarding the labeling of news sources and SI, Table S5, for a table showing the percentage of domains matched with a MBFC rating).

\begin{figure*}[tbp]
  \centering
  \includegraphics[width=\columnwidth]{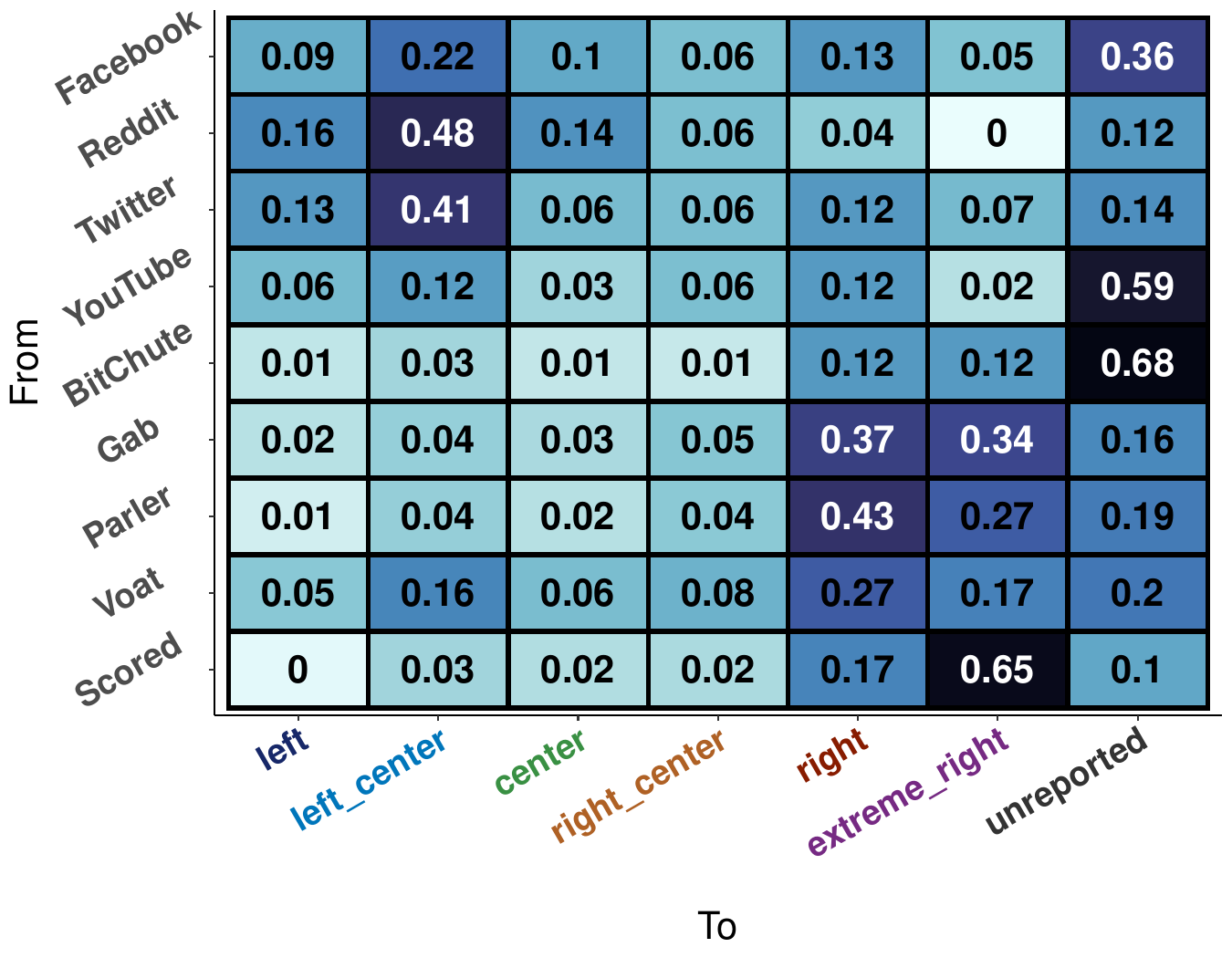} 
  \caption{\textbf{News Consumption}.
   Fraction of URLs directing to domains with differing political leanings, as categorized by MBFC.
   Domains without a political bias identified by MBFC are categorized as ``unreported''. 
   News consumption on mainstream platforms is skewed toward left-leaning content, while alt-tech platforms predominantly link to right-leaning content. 
   YouTube and BitChute prominently share ``unreported'' domains.
   }
  \label{fig:diet}
\end{figure*}

Fig.~\ref{fig:diet} illustrates users' news consumption across platforms, showing the proportion of URLs directing to domains with differing political leanings, as categorized by MBFC.
A clear pattern emerges: mainstream platforms such as Facebook, Reddit, and Twitter tend to skew toward left-leaning content, while alt-tech platforms—including Gab, Parler, Voat, and Scored—predominantly link to right-leaning content. 
Voat stands out among alt-tech platforms by directing a notable portion of its traffic to left- or left-center-leaning domains. However, 89.5\% of these links were shared across 6 of the 60 distinct \textit{subverses} analyzed, most of which (e.g., v/AnonAll, v/GreatAwakening) align with alt-right or conspiracy groups (a detailed list of these \textit{subverses} is available in Table S4 of the SI). 
This pattern suggests that these links may have been shared in an ironic or derogatory manner. 
YouTube and BitChute, despite being filtered for political content, heavily link to ``unreported" domains, reflecting their role as platforms that host a broad range of less conventional or unvetted content. 
Additionally, on BitChute, a substantial share of content from hyper-extremist sites is driven by a small group of highly active users. 
In some cases, these users are only a few dozen or hundreds, yet they are responsible for thousands of links, creating a disproportionate volume of extremist content on the platform. 
This finding aligns with the concept of ``vocal minorities''~\cite{mustafaraj2011vocal}, where a small but highly active subset of users exerts an outsized influence on the platform by driving a significant share of its activity~\cite{nogara2022disinformation}.

To further illustrate these patterns, Table~\ref{table:hubs} reports each platform's ten most frequently linked domains. While Facebook and Twitter often feature sources with a left-center bias, such as CNN, The New York Times, or The Washington Post, Reddit displays a more homogenous profile as no right-biased sources appear among the most linked domains. In contrast, platforms such as Gab, Parler, Scored, and Voat direct a substantial proportion of their traffic to sources that are both right-biased and have low reliability. 
Scored, for instance, channels 58\% of its external traffic to ``Patriots.Win", an online forum created following the ban of the ``r/The\_Donald" subreddit in June 2015~\cite{horta2021platform}. This forum is known for using unreliable sources and a lack of transparency and moderation \cite{mbfc_patriotswin, adl_patriotswin}. Although both YouTube and BitChute frequently link to unreported domains, BitChute's links often direct users to alternative news websites, including domains associated with the american alt-right.

\begin{table*}[t]
    \centering
    \scriptsize
    \caption{Most shared domains by platform. Name and percentage of links toward the ten domains most linked by each platform. Each domain is color coded to represent its political bias as follows: \textit{\textbf{\textcolor{left}{left}}, \textbf{\textcolor{left-center}{left-center}}, \textbf{\textcolor{center}{center}}, \textbf{\textcolor{right-center}{right-center}}, \textbf{\textcolor{right}{right}}, \textbf{\textcolor{extreme-right}{extreme-right}}, \textbf{unreported}}. An asterisk following a domain's name indicates that it is flagged as a questionable source.}
     \begin{adjustbox}{max width=\textwidth}
    \begin{tabular}{lcc|cc|cc|cc|cc}
        \toprule
        \multirow{2}{*}{} & \multicolumn{2}{c}{Facebook} & \multicolumn{2}{c}{Reddit} & \multicolumn{2}{c}{Twitter} & \multicolumn{2}{c}{YouTube} & \multicolumn{2}{c}{BitChute} \\
        \cmidrule(lr){2-3} \cmidrule(lr){4-5} \cmidrule(lr){6-7} \cmidrule(lr){8-9} \cmidrule(lr){10-11}
        & Domain & \% links & Domain & \% links & Domain & \% links & Domain & \% links & Domain & \% links \\
        \midrule
        & \textcolor{left-center}{CNN} & 2.35  & \textcolor{center}{The Hill} & 6.00 & \textcolor{left-center}{The New York Times}  & 6.90 & \textcolor{right}{Fox News}*  & 8.09 & \textcolor{extreme-right}{Infowars}* & 3.14 \\ 
        & \textcolor{right}{Fox News}*  & 2.30  & \textcolor{left-center}{CNN} & 5.82 & \textcolor{left-center}{The Washington Post} & 6.56 & \textcolor{left}{MSNBC}  & 1.97 & Liberty Classroom* & 2.27 \\ 
        & \textcolor{center}{The Hill} & 1.79  & \textcolor{left-center}{The Washington Post} & 4.73 & \textcolor{left-center}{CNN} & 5.72 & {Tv9Hindi} & 1.86 & \textcolor{right}{Rebel News}* & 1.65 \\
        & \textcolor{left-center}{The Washington Post} & 1.72  & \textcolor{left-center}{The New York Times}  & 3.95 & \textcolor{right}{Fox News}* & 3.22 & \textcolor{left}{The Young Turks} & 1.77 & {Tuttle Twins} & 1.65 \\ 
        & \textcolor{left-center}{The New York Times}  & 1.65  & \textcolor{left-center}{Politico} & 2.79 & \textcolor{left-center}{Politico} & 2.34 & \textcolor{left-center}{CNBC} & 1.55 & \textcolor{extreme-right}{The Gateway Pundit}* & 1.56 \\ 
        & \textcolor{left-center}{NBC News}    & 1.57  & \textcolor{center}{Reuters} & 2.37 & \textcolor{center}{The Hill}  & 2.30 & \textcolor{left-center}{PBS NewsHour} & 1.47 & \textcolor{right}{TimCast} & 1.53 \\ 
        & \textcolor{left-center}{Yahoo News}  & 1.50  & \textcolor{left-center}{Associated Press}  & 2.18 & \textcolor{left-center}{NBC News}  & 2.27 & \textcolor{right-center}{Fox Business} & 1.47 & {Martin Brodel 1776}  & 1.51  \\ 
        & \textcolor{right}{Breitbart}*    & 1.46   & \textcolor{left-center}{The Guardian} & 2.10 & \textcolor{right}{Breitbart}*  & 2.23 & \textcolor{left-center}{NBC News} & 1.10 & PeteLive  & 1.48 \\ 
        & \textcolor{right}{Daily Wire}   & 1.22       & \textcolor{left-center}{NBC News} & 2.07 & \textcolor{left}{Raw Story}  & 2.19 & France 24      & 0.91 &\textcolor{extreme-right}{TurleyTalks}*  & 1.45 \\ 
        & \textcolor{left-center}{MSN}  & 1.04  & \textcolor{left-center}{Business Insider} & 1.93 & \textcolor{extreme-right}{The Gateway Pundit}* & 1.92 & \textcolor{left-center}{CBS News}        & 0.67 & \textcolor{right}{Banned}*  & 1.17 \\ 
        \midrule

       \multirow{2}{*}{} & \multicolumn{2}{c}{Gab} & \multicolumn{2}{c}{Parler} & \multicolumn{2}{c}{Scored} & \multicolumn{2}{c}{Voat} & {} \\ 
\cmidrule(lr){2-3} \cmidrule(lr){4-5} \cmidrule(lr){6-7} \cmidrule(lr){8-9}  
& Domain & \% links & Domain & \% links & Domain & \% links & Domain & \% links & &  \\ 
\cmidrule(lr){2-9} 
& \textcolor{extreme-right}{The Gateway Pundit}* & 13.71 & \textcolor{extreme-right}{The Gateway Pundit}* & 13.74 & \textcolor{extreme-right}{Patriots}* & 58.24 & \textcolor{extreme-right}{The Gateway Pundit}* & 7.47 & \\ 
& \textcolor{right}{Breitbart}* & 12.07 & \textcolor{right}{Fox News}* & 11.44 & \textcolor{right}{Breitbart}* & 3.39 & \textcolor{right}{Breitbart}* & 5.58 & & \cellcolor{left}\textcolor{white}{Left}  \\ 
& \textcolor{right}{Fox News}* & 4.09 & \textcolor{right}{Breitbart}* & 8.56 & \textcolor{right}{MAGA}* & 2.64 & \textcolor{right}{Zero Hedge}* & 4.91 &  & \cellcolor{left-center}\textcolor{white}{Left Center}\\ 
& \textcolor{right}{Zero Hedge}* & 2.47 & \textcolor{right}{The Epoch Times}* & 4.94 & \textcolor{extreme-right}{The Gateway Pundit}* & 2.33 & \textcolor{right}{Fox News}* & 2.49 & &\cellcolor{center}\textcolor{white}{Center} \\ 
& \textcolor{right}{The Epoch Times}* & 2.43 & \textcolor{extreme-right}{Western Journal}* & 2.44 & \textcolor{right}{Fox News}* & 1.65 & \textcolor{right}{DailyMail}* & 1.92 & &\cellcolor{right-center}\textcolor{white}{Right Center} \\ 
& \textcolor{extreme-right}{InfoWars}* & 2.29 & \textcolor{right-center}{New York Post} & 1.74 & \textcolor{right}{Zero Hedge}* & 0.99 & \textcolor{center}{The Hill} & 1.91 & & \cellcolor{right}\textcolor{white}{Right} \\ 
& \textcolor{extreme-right}{GNews}* & 1.97 & \textcolor{right}{NewsMax}* & 1.32 & \textcolor{right-center}{New York Post} & 0.92 & \textcolor{right-center}{RT News}* & 1.73 & &\cellcolor{extreme-right}\textcolor{white}{Extreme Right}  \\ 
& \textcolor{right-center}{New York Post} & 1.70 & \textcolor{right}{Town Hall}* & 1.31 & \textcolor{right}{Washington Examiner}* & 0.73 & \textcolor{right-center}{New York Post} & 1.57 & &\cellcolor{black}\textcolor{white}{Unreported} \\ 
& \textcolor{center}{The Hill} & 1.31 & \textcolor{extreme-right}{Just The News}* & 1.26 & \textcolor{right}{DailyMail}* & 0.70 & \textcolor{left-center}{CNN} & 1.44 &   \\ 
& \textcolor{right}{DailyCaller} & 1.19 & {CounterGlobalist} & 1.16 & \textcolor{right}{DailyCaller} & 0.56 & \textcolor{right}{Washington Examiner} & 1.35 &  \\ 

        \cmidrule(lr){2-9}         
        \bottomrule
    \end{tabular}
    \end{adjustbox}
    \label{table:hubs}
\end{table*}

Next, we analyze the similarity between the news diets across platforms. 
To this aim, we identify the top 20 most-linked domains for each platform and compute the weighted cosine similarity between the domain vectors for all pairs.  
This method quantitatively measures how closely platforms align in their shared content. 

\begin{figure}[h]
  \centering
  \includegraphics[width=\columnwidth]{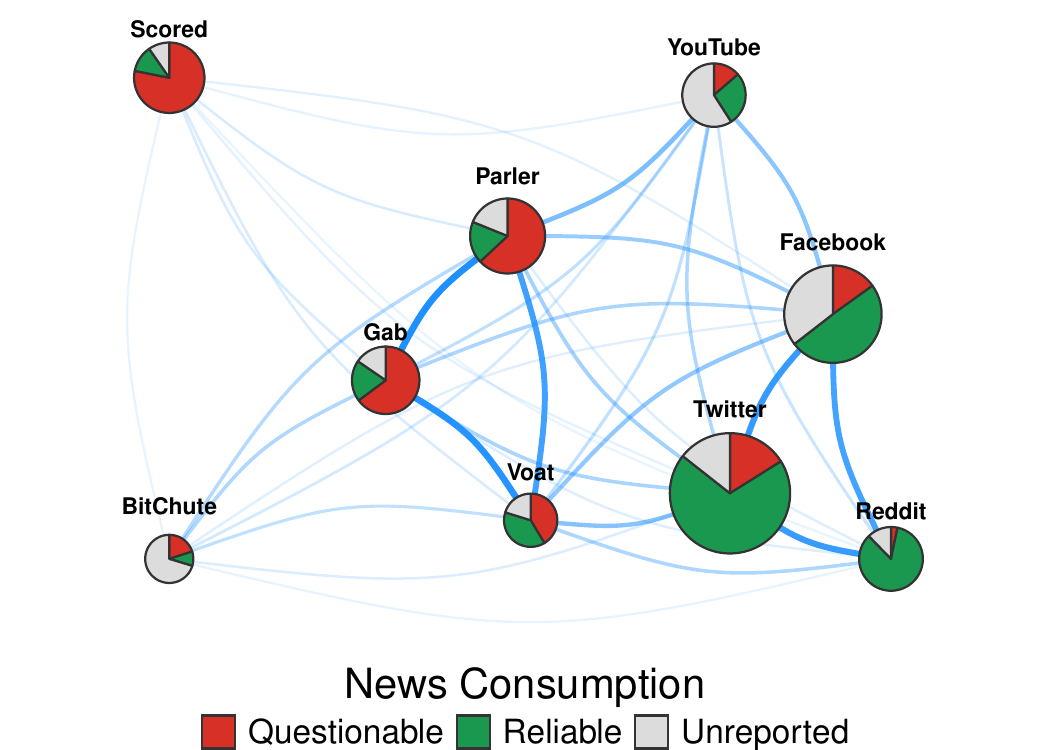} 
  \caption{\textbf{Platform similarity.} Cosine similarity network based on the platforms' 20 most linked domains. 
  The size of the different nodes is proportional to the volume of links shared in the platform, while the colors of the pies indicate the fraction of questionable or reliable content shared. 
  We observe two cliques with high similarity: one made up of mainstream platforms (Facebook, Twitter, Reddit) that share a majority of reliable news sources, and one made up of alt-tech ones (Gab, Parler, Voat) sharing a higher fraction of questionable sources. Scored, BitChute, and, to an extent, YouTube remain fairly separated from the rest of the platforms.}
  \label{fig:cosine}
\end{figure}

The results, illustrated in Fig.~\ref{fig:cosine}, reveal two distinct cliques characterized by high similarity scores. 
The first clique consists of the mainstream platforms Facebook, Twitter, and Reddit, all exhibiting cosine similarity scores above $0.7$. 
The second clique includes the alt-tech platforms Gab, Voat, and Parler, with similarity scores ranging from $0.73$ to $0.88$. See SI for the full cosine similarity matrix and for a robustness check expanding the threshold of the top 20 most-linked domains (Fig. S2, S3).

The similarity between mainstream and alt-tech platforms generally does not exceed $0.5$, with Facebook and Twitter showing slightly higher similarity to specific fringe platforms. 
This last observation may indicate that news consumption withing certain echo chambers on Facebook and Twitter partially overlaps with those on fringe platforms.
Reddit appears to be the mainstream platform most dissimilar to alt-tech platforms, exhibiting the lowest similarity in news consumption. 
This aligns with the tendency of Reddit users to predominantly share left-leaning content, while extreme-right sources are absent and right-center or right-leaning content is relatively uncommon.
YouTube shows moderate similarity to other mainstream platforms, particularly Twitter, but also with some alt-tech ones, such as Parler.
Finally, BitChute and Scored are dissimilar to all other platforms. 
This difference may be attributed to the extreme nature of the content frequently shared on BitChute and the limited popularity and reach of Scored, which results in unique news diets that diverge significantly both from mainstream and other alt-tech platforms.

Fig.~\ref{fig:cosine} also highlights the fraction of questionable domains (as defined by MBFC) shared by each platform. 
Mainstream platforms share a relatively small fraction of unreliable sources, while alt-tech platforms display significantly higher proportions of such content, reinforcing the distinct content dynamics within the social media ecosystem.
These results are reported in more detail in Table~\ref{tab:datasetbreakdown}, showing the fraction of questionable sources shared by users on each platform. 
Alt-tech platforms Gab, Parler, and especially Scored exhibit the highest proportions. 
Voat shows a lower fraction, and Reddit stands out for its smallest fraction of questionable sources shared ($0.03$). 
These findings underscore the distinction between alt-tech platforms, where questionable sources appear to circulate more freely, and mainstream platforms, where such content constitutes only a small proportion of shared material.
BitChute shows a small value of $0.22$, differing from the general trend observed among alt-tech platforms, but aligning with the platform's propensity to host a significant volume of unreported domains.

\subsection*{\textbf{User base}}

To assess the diversity of the user bases across platforms, we evaluate the distribution of users' political leaning.
We infer the political leaning of active users (i.e., users who shared 10 or more URLs toward domains with an associated political bias) based on their posting activity. 
We assign a numerical score to news media, ranging from $-1$ (extreme-left) to $+1$ (extreme-right), excluding unreported sources. The political leaning of each user is then defined as the average of the leanings of the news domains they post (see section Materials and Methods for further details). 
We note that YouTube and BitChute operate on different principles compared to the other platforms considered. As video-sharing platforms, they lack traditional feeds or direct user interactions. Consequently, our estimation of political leanings on these platforms reflects the tendencies of active content producers rather than user consumption patterns. 
Additionally, we acknowledge that using the average as a summary statistic may fail to capture behaviors more evenly distributed across a bias spectrum: as such, a robustness check showing the distribution of the variance of the political leaning scores assigned to users is available in SI (Fig. S4). The analysis of the variance at the user level shows that most of them tend to engage with content from a narrow ideological range, making the average a quite representative summary statistic of the general trend.

\begin{figure*}[t]
  \centering
  \includegraphics[width=\textwidth]{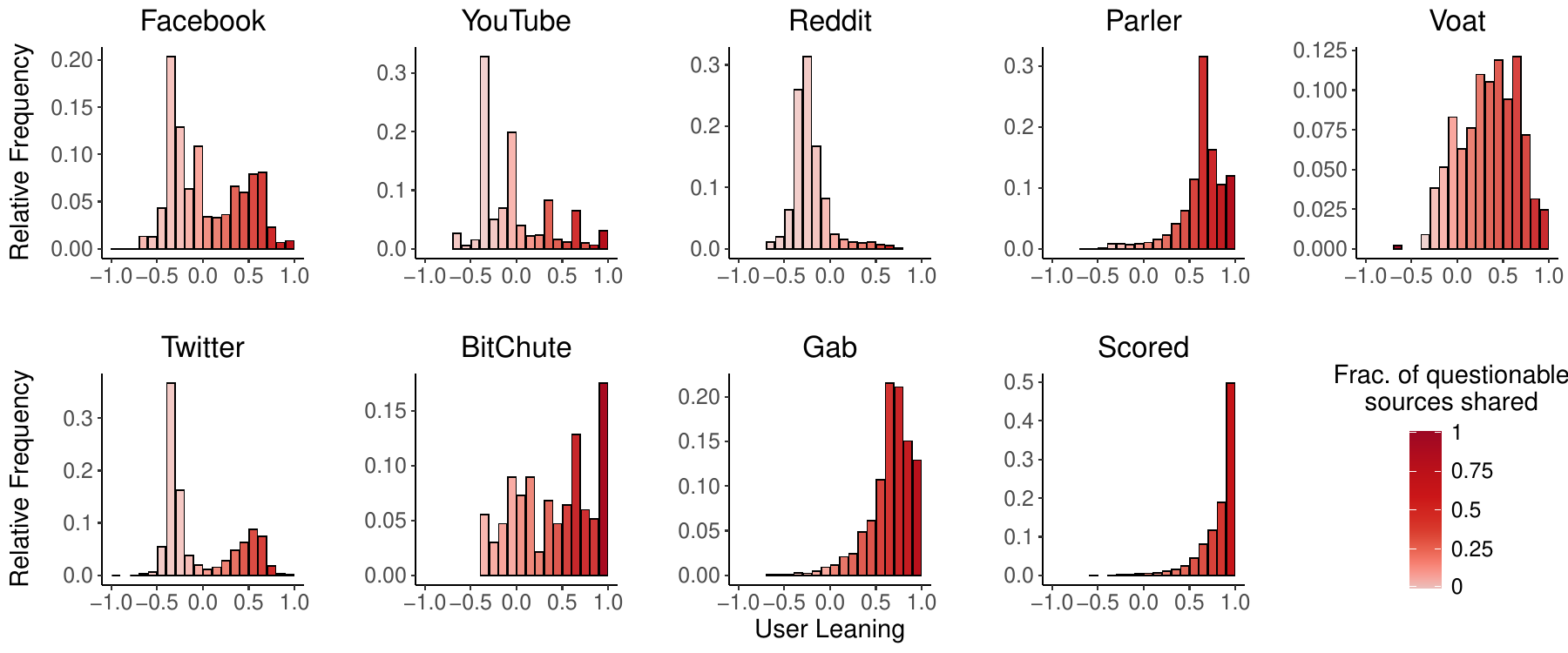} 
  \caption{\textbf{User base composition.} Distributions of users' political leaning for each platform. Each unique user gets assigned a leaning score between $-1$ and $+1$, according to their posting activity. The bars are colored according to the number of ``questionable" sources shared by users of a specific leaning. We notice how Facebook and Twitter have a polarized user base with two distinct groups, one sharing mostly reliable content and the other sharing mostly content with low factual reporting. The situation is more homogeneous regarding all of the fringe platforms, and the remaining two mainstream platforms.}
  \label{fig:user_leaning}
\end{figure*}

Fig.~\ref{fig:user_leaning} illustrates the political leaning distribution of users in the nine platforms. 
We note that the user bases of Facebook and Twitter are made up of two distinct groups that span the entire ideological spectrum, resulting in a bimodal distribution of political leaning. 
The same pattern is not observed on some alt-tech platforms (Gab, Parler, and Scored), where users are concentrated within a narrower leaning interval, forming a homogeneous, right-leaning user base. 
Interestingly, Reddit exhibits a skewed leaning distribution, similar to alt-tech platforms, but concentrated on left leaning content, as noted in previous work~\cite{de2021no, monti2023evidence}. Voat and BitChute display a slightly more heterogeneous user base compared to other alt-tech platforms. 
For Voat, this heterogeneity likely stems from the significant portion of traffic redirected toward left-biased sources by alt-right groups, as previously observed. 
For BitChute, the high proportion of unreported sources on the platform likely contributes to its heterogeneity. Specifically, since we only include users who shared at least 10 URLs with an associated political bias in our computation, this criterion captures only a small subset of the user base.

Expanding on BitChute, users with a political leaning below 0 (left-leaning) make up approximately 12\% of the total user base but share, on average, 15 fewer links than right-leaning users (leaning above 0) and 30 fewer links than those with a leaning equal or above $0.5$. 
This suggests that left-leaning users are generally less active in sharing content, resulting in a smaller contribution to the circulation of news on the platform. 
These dynamics underscore an asymmetry in content production and dissemination across ideological groups, highlighting the influence of vocal communities in shaping platform-wide narratives.

The observations regarding the diversity of the user bases can be quantified by means of the variance of the political leaning distribution, $\sigma^2$, reported in Table~\ref{tab:datasetbreakdown}. 
Platforms such as Twitter, Facebook, YouTube, BitChute, and Voat to some extent, exhibit higher variance, indicating wide spread user leanings and reflecting a diverse user base, while the remaining platforms demonstrate more homogeneous user bases.
When considering the reliability of the sources shared, we observe that users with right-leaning and extreme-right political orientations share a higher proportion of questionable content. However, this pattern aligns with established correlations reported by MBFC and is not specific to our study.

\section*{Conclusions}

This study characterized fragmentation in the social media ecosystem related to the 2020 US presidential election, and introduced the concept of ``echo platforms'', i.e., entire social media platforms that operate as self-contained echo chambers, reinforcing homogeneous beliefs and isolating users from opposing viewpoints. 
Echo platforms represent an evolution of traditional echo chambers, where the platform structure fosters ideological uniformity. 
Our framework categorizes platforms across three dimensions—platform centrality, content reliability, and user base homogeneity—providing a systematic approach to understanding these dynamics within the broader context of social media fragmentation.

Such analysis provides a better characterization of the differences between mainstream platforms (e.g., Facebook, Twitter) and alt-tech ones (e.g., Gab, Parler).
In particular, we uncover that Reddit shares some features with mainstream platforms, like the high reliability of posted news media, but others with fringe platforms, such as a uniform user composition and a relatively peripheral role in cross-platform linking. 
Our findings suggest that, despite being a mainstream social media platform, Reddit functions more like an echo-platform than others.

This systemic shift from segregated communities to platform-wide segregation highlights the role of moderation policies, user self-selection, and engagement-driven business models in shaping the rise of echo platforms.
Our findings contribute to shedding light on the factors influencing digital polarization, showing how alt-tech platforms attract marginalized or de-platformed communities from mainstream spaces, creating distinct niches that intensify ideological separation. 
This underscores the need to reevaluate moderation strategies, as current policies may unintentionally exacerbate the fragmentation of the social media ecosystem by fostering the growth of insulated echo platforms. 
Such dynamics challenge the digital public sphere by limiting constructive dialogue and deepening ideological divides.

While our framework provides robust tools for identifying and analyzing echo platforms, the study has limitations. 
The dataset, though extensive, reflects a specific period marked by significant political events, such as elections.
This context offers valuable insights into platform behavior during critical societal moments but may only partially capture ongoing changes in the social media landscape. 
Future research could extend this work by integrating more recent data and exploring cross-cultural or longitudinal variations in platform dynamics, as well as investigating the impact of coordinated actions and in the fragmentation of online spaces, and, where feasible, incorporating cross-platform user identification to enhance the depth of analysis.

Our results offer a foundation for examining the socio-political impacts of echo platforms, particularly in contexts where public discourse and social cohesion are at risk. Understanding and mitigating the societal effects of echo platforms is essential for developing policies that balance open expression with fostering a diverse and cohesive public sphere.

This study highlights the influence of echo platforms in shaping today's social media landscape, particularly their role in internet fragmentation and societal polarization.
By offering a detailed quantitative approach, we aim to advance understanding of these platforms and inform strategies to mitigate the adverse impacts of digital fragmentation on society.

\section*{Methods}
\subsection*{\textbf{Data}}
In this section, we describe the procedures for data collection and preprocessing, along with details about the datasets used in our study. All of the datasets were filtered to retrieve politically relevant content pertaining to the 2020 U.S. election, either by performing a keyword search, or by selecting relevant political communities where a keyword search was not feasible. The set of keywords is consistent among the datasets in which they were utilized (with minor variations), and based on the presidential candidates' names.
During data preprocessing, we manually excluded URLs that were not relevant to our analysis. This included self-links (i.e., a platform linking to itself), URLs directing to social media platforms not included in the datasets (e.g., Snapchat, TikTok), video streaming services (e.g., Vimeo, Dailymotion), financial services (e.g., PayPal, Venmo), music streaming services (e.g., Spotify, SoundCloud), and tech platforms or services (e.g., Google Suite, Streamlabs). Additionally, we removed a range of URLs that did not fit these broad categories but were similarly uninformative for our purposes, such as links to Amazon, Steam, Coinbase, and NASA.\\

\textbf{Facebook:} The URLs analyzed were extracted from 21 million Facebook posts collected using the CrowdTangle service. These posts were identified based on searches using predefined keyword lists: L1: \{Trump, trump, \#donaldtrump, \#trump\} and L2: \{Biden, biden, \#joebiden, \#biden\}, spanning the period from May to November 2020.\\

\textbf{Reddit:} We utilize URLs extracted from a comprehensive collection of posts published in the subreddit r/Politics between January and December 2020, totaling nearly 4.7 million posts. These posts were collected via the Pushshift dataset~\cite{baumgartner2020pushshift}. A subreddit is a user-created community focused on specific topics and governed by its own rules, where members can subscribe, share posts, comment, and upvote or downvote content.\\

\textbf{Twitter:} The dataset~\cite{dataflamino2023political}, based on information retrieved from Flamino~\emph{et al.}~\cite{flamino2023political}, was obtained using the Twitter Search API, with the names of the two U.S. presidential candidates from the 2020 election as keywords (consistently with Facebook). It includes 174 million tweets posted between June and November 2020.\\

\textbf{YouTube:} We use URLs extracted from the descriptions of 270,000 YouTube videos, collected using the YouTube Data API between June and December 2020. The videos were identified through searches based on the same keywords utilized for Facebook. For each video, an additional search was conducted by crawling the network of related videos, as suggested by YouTube's algorithm. From the gathered dataset, we retained only videos containing {$Trump$|$trump$} (or {$Biden$|$biden$}, respectively) in the title or description.
\\

\textbf{BitChute:} BitChute is a British alt-tech video hosting platform known for its lower moderation efforts compared to its more popular counterpart, YouTube. The platform focuses heavily on news and politics and is associated with a significant amount of hate speech in both videos and comment sections~\cite{trujillo2004bitchute}. For our analysis, we use the MeLa BitChute Dataset~\cite{DVN/KRD1VS_2022}, which provides a near-complete scrape of the platform from 2019 to 2021, from which we filtered the data to retain only the 1.1 million videos uploaded in 2020. We further refined the dataset by performing a keyword search on the video titles. The keywords used for filtering were \{trump, biden, joebiden, donaldtrump, donald, trump2020, biden2020\}, resulting in the retention of 40 thousand videos.\\

\textbf{Gab:} Gab is an alt-tech microblogging platform structured similarly to Twitter but with minimal content moderation, promoting free speech and Christian values~\cite{wpost2018gab, Forward2022}. With its predominantly far-right user base, Gab has been described as a safe haven for neo-Nazis, members of the American alt-right, Trump supporters, and conspiracy theorists~\cite{atlantic2018gab, foxnews2020gab}. The platform has also been repeatedly linked to online radicalization and real-world violent events~\cite{ribeiro2021evolution, levenson2018gab}.

The data for this study was collected in two phases. From June 1, 2020, to October 23, 2020, posts were downloaded using Gab's general stream, capturing all posts generated during this period. After October 23, due to the deprecation of the API endpoint, data was collected using the timelines of 930,000 users identified in the first phase. The dataset was then filtered using the keywords \{``trump", ``biden"\}, resulting in 467,000 posts made between June 1 and December 1, 2020.\\

\textbf{Parler:} Launched in 2018, Parler is a platform with functionalities similar to Twitter, marketed as a free-speech-focused alternative. It gained mainstream attention following the storming of Capitol Hill in 2021, as it was one of the platforms allegedly used to incite and plan the attack~\cite{GaisCruz2021, ABCNews2021}. Shortly after, Parler was removed from the Google Play Store and Apple App Store and subsequently suspended by its hosting provider, Amazon AWS. At the time of writing, plans to relaunch Parler's services have been announced, though the exact timeline remains unclear.

The data used in this study comes from a large dataset comprising 183 million posts made between 2018 and 2021, collected by Aliapoulios~\emph{et al.}~\cite{aliapoulios2021large}. For our analysis, we focused exclusively on the 1.7 million posts made during 2020 containing at least one of the keywords \{trump, biden, joebiden, donaldtrump, donald, trump2020, biden2020\} in their main body.\\

\textbf{Scored:} Scored (accessible via both \url{https://scored.co} and \url{https://communities.win}) emerged as an alternative to Reddit, hosting numerous communities that were banned from more prominent social networks, including $c/TheDonald$, $c/GreatAwakening$, and $c/FatPeopleHate$. For this study, we use the iDRAMA-Scored-2024 dataset~\cite{patel2024idrama}, a comprehensive scrape of the platform since its inception in 2020, for a total of 6.2 million posts spanning nearly four years. We filter the dataset by retaining only the submissions posted during 2020 in six politically centered communities of the 975 present communities, which account for 92\% of links shared during 2020 over 1.4M posts. Please refer to the SI for a detailed table of the selected communities retained in the analysis (Table S3).\\

\textbf{Voat:} The now-defunct Voat, which was shut down in December 2020, served as an alternative to Reddit, similar to Scored. Several communities banned from Reddit migrated to Voat, forming new \textit{subverses}—the platform's equivalent of subreddits. Notable examples include v/fatpeoplehate, v/TheRedPill, and v/GreatAwakening. For this study, we use data from a collection of 2.3 million posts made on the platform between November 2013 and December 2020, compiled by Mekacher and Papasavva~\cite{mekacher2022can}. From this dataset, we retain only submissions made during 2020 across the 60 most prominent political subvoats (over the $7,604$ total subvoats), for a total of almost 200 thousands posts. Please refer to the SI for a table of the selected subverses retained in our analysis (Table S4).

\subsection*{\textbf{Platforms graph and null model}}
The set of connections (URLs) between platforms can be modeled as a weighted directed graph $ G = (V, E, w)$, where $V$ represents the set of nine social media platforms, $E \subseteq V \times V$ the set of directed edges between platforms, and $w: E \to \mathbb{R}^+$ the weight of edges corresponding to the number of times one platform links to another. Given the size disparity between the considered data sets, quantifying the relevance of links among nodes using only their weights would introduce a bias. Thus, we compute how such weights deviate from their expected values. These expected values can be computed using the weighted configuration model~\cite{serrano2005weighted}, a null model in which the in- and out-strength distributions of the nodes are preserved. 
According to the weighted configuration model, the average weight of a link connecting two uncorrelated vertices with out-strength $s^{out}_i$ and in-strength $s^{in}_j$ can be written as $\mathbb{E}[w_{ij}] = \frac{{s^{out}_i \cdot s^{in}_j}}{S}$, where $S$ refers to the total weight of the network. Hence, we can compute the relationship between the observed weights and their expectation as follows:  
\begin{equation*}
\hat{w}_{i,j} = \frac{w_{i,j}}{\mathbb{E}[w_{ij}]} \,.
\end{equation*}
Such quantity approximates the results we would obtain by performing a strength-preserving randomization on the network. If higher than 1, the weights we obtain indicate that a link from one platform to another occurs more often than we would expect at random, and vice versa.

\subsection*{\textbf{Labeling of news sources}}
As mentioned, we utilize Media Bias/Fact Check (MBFC, https://mediabiasfactcheck.com/) to label news outlets based on their political bias and level of reliability regarding factuality reporting. MBFC is an independent fact-checking organization that rates various news sources. The labeling utilized in this study, collected in October 2024, contains political bias categories ranging from extreme-left to extreme-right, while certain news outlets are classified as ``questionable", indicating that they exhibit one or more of the following, per MBFC: ``extreme bias, consistent promotion of propaganda/conspiracies, poor or no sourcing to credible information, a complete lack of transparency and/or is fake news''. While not explicitly falling under the ``questionable'' definition of MBFC, we also consider as such domains classified as ``conspiracy/pseudoscience'', given their inherent low credibility and absence of proper fact-checking. Furthermore, we manually labeled a small number of right-biased and/or extremist websites which accounted for a significant portion of some alt-tech platforms' traffic, but that were not present on MBFC.

\subsection*{\textbf{Inferring accounts' leaning}}
To compute an account's leaning, we utilize the following algorithmic procedure: we assign a score between $-1$ and $+1$ to each external domain, depending on its MBFC's political bias label. Namely, -1 for the extreme left, -0.66 for left, -0.33 for left-center, 0 for least biased, 0.33 for right-center, 0.66 for right, and +1 for the extreme right. For an account $i$  who shared $n$ URLs towards external domains $C_i = \{c_1, c_2,...,c_n\}$, each URL $c_j$ is associated with one of these numeric values. The political leaning $x_i$ of the account $i$ is then defined as the average of the political bias of all the domains shared: 
\begin{equation*} 
x_i \equiv \frac{\sum^{n}_{j=1}c_j}{n}.
\end{equation*}
This returns a leaning score in the interval $[-1, 1]$, where a value of $-1$ ($+1$) indicate an extreme-left (extreme-right) leaning. This procedure, though simple, is grounded in psychological theories such as selective exposure~\cite{stroud2010polarization} and has proven to be an effective estimator of users' political leaning \cite{cinelli2021echo,flamino2023political}.

\clearpage

\newcommand{\beginsupplement}{
    \setcounter{section}{0}
    \renewcommand{\thesection}{S\arabic{section}}
    \setcounter{equation}{0}
    \renewcommand{\theequation}{S\arabic{equation}}
    \setcounter{table}{0}
    \renewcommand{\thetable}{S\arabic{table}}
    \setcounter{figure}{0}
    \renewcommand{\thefigure}{S\arabic{figure}}
    \newcounter{SIfig}
    \renewcommand{\theSIfig}{S\arabic{SIfig}}}

\beginsupplement

\onecolumn
\section*{\centering Supplementary Information}

\section*{SI Appendix Section}
\label{sec:sample:appendix}

\begin{table*}[h]
    \centering
    \renewcommand{\arraystretch}{1.5}
     \setlength{\tabcolsep}{0.9\tabcolsep}
    \caption{Timeframe and number of URLs collected (before and after processing) for each platform's data set.}
   \begin{tabular}{|l|c|c|c|c|c|c|c|c|}
        \hline
        \textbf{Platform} & Timeframe & \# of URLs (unprocessed) & \# of URLs (processed)   \\
        \hline
        Facebook   & 25/05/2020 - 15/11/2020 & 20M & 10.7M  \\
        \hline
        Reddit     & 01/01/2020 - 31/12/2020 & 328k & 320k  \\
        \hline
        Twitter    & 01/06/2020 - 03/11/2020 & 189M & 103M   \\
        \hline
        YouTube    & 01/06/2020 - 16/12/2020 & 981k & 510k  \\
        \hline
        BitChute   & 02/04/2020 - 09/10/2020 & 3.2M & 107k  \\
        \hline
        Gab        & 01/06/2020 - 01/12/2020 & 468k & 453k  \\
        \hline
        Parler     & 01/01/2020 - 31/12/2020 & 14M & 1.1M   \\
        \hline
        Scored     & 01/01/2020 - 31/12/2020 & 984k & 868k  \\
        \hline
        Voat       & 01/01/2020 - 25/12/2020 & 269k & 135k \\
        \hline
    \end{tabular}
    \label{tab:timeframe}
\end{table*}

\begin{table*}[h]
\centering
\caption{Table showing the subverse name, number of URLs shared, number of URLs shared toward left biased sources, fraction of left biased sources shared, cumulative count and cumulative proportion of left biased sources shared for the six Voat's subverses responsible for sharing 89.5\% of the platform's left biased content. Note how in this table we consider as ``left biased'' every source classified as either extreme-left biased, left biased, or left-center biased.}
\begin{tabular}{lccccc}
Subverse & URLs  & left-leaning URLs  & frac. left  & cum. count  & cum. prop. \\
\midrule
1. AnonAll & 15822 & 9643 & $0.609$ & 9643 & $0.399$\\
2. news & 31884 & 5931 & $0.186$ & 15574 & $0.644$\\
3. politics & 20898 & 2627 & $0.125$ & 18201 & $0.753$\\
4. GreatAwakening & 35987 & 1535 & $0.042$ & 19736 & $0.816$\\
5. OccidentalEnclave & 4865 & 1345 & $0.276$ & 21081 & $0.872$\\
6. Niggers & 5621 & 553 & $0.098$ & 21634 & $0.895$\\
\bottomrule
\end{tabular}
\end{table*}

\begin{table*}[h]
\caption{Selected Scored Communities retained in our analysis, and number of URLs contained in them} 
\label{tab:selected_communities} 
    \centering
    \begin{tabular}{lcr}
        \toprule
        \# & Community & Nr. of URLs\\
        \midrule
        1 & TheDonald  & 907439\\
        2 & OmegaCanada & 7034\\
        3 & GreatAwakening & 2804\\
        4 & GavinMcInnes & 1248\\
        5 & Conspiracies & 1180\\
        6 & Conservative & 224\\
        \bottomrule
    \end{tabular}
\end{table*}

\clearpage

\begin{longtable}{lcr}
\caption{\textbf{Selected Subverses retained in our analysis, and number of URLs contained in them.}} \label{tab:selected_subvoat} \\
\toprule
\# & Subverse name & Nr. of URLs \\
\midrule
\endfirsthead

\multicolumn{3}{c}%
{{\bfseries \tablename\ \thetable{} -- continued from previous page}} \\
\toprule
\# & Subverse name & Nr. of URLs \\
\midrule
\endhead

\midrule \multicolumn{3}{r}{{Continued on next page}} \\
\endfoot

\bottomrule
\endlastfoot

1 & GreatAwakening & 35987\\
2 & news & 31884\\
3 & politics & 20898\\
4 & AnonAll & 15822\\
5 & theawakening & 12765\\
6 & Niggers & 5621\\
7 & OccidentalEnclave & 4865\\
8 & Conspiracy & 3766\\
9 & WorldToday & 2753\\
10 & Worldnews & 2289\\
11 & TheDonald & 1620\\
12 & TheGreatAwakening & 920\\
13 & USNews & 776\\
14 & politicalnews & 754\\
15 & 2020ElectionNews & 749\\
16 & Worldpolitics & 693\\
17 & uspolitics & 451\\
18 & CRIMENEWS & 431\\
19 & Jews & 359\\
20 & christiannews & 284\\
21 & ChristianEnclave & 281\\
22 & CultureWars & 278\\
23 & economics & 250\\
24 & SJWHate & 221\\
25 & PivottoAsia & 219\\
26 & AnonNews & 215\\
27 & Libertarian & 161\\
28 & pizzagate & 160\\
29 & CanadaFirst & 150\\
30 & Military & 145\\
31 & pedogate & 140\\
32 & PoliticalDiscussion & 123\\
33 & BadCopNoDonut & 112\\
34 & pizzagateuncensored & 112\\
35 & education & 111\\
36 & HillaryforJail & 76\\
37 & Lawenforcement & 73\\
38 & PoliticallyIncorrect & 69\\
39 & BlackLivesMatter & 68\\
40 & conspiracyfact & 66\\
41 & newsandpolitics & 61\\
42 & PedogateFullExposure & 59\\
43 & WhiteRights & 58\\
44 & ANTIFAWATCH & 55\\
45 & gunpolitics & 54\\
46 & Immigration & 54\\
47 & Identitarians & 47\\
48 & ClimateSkeptics & 44\\
49 & ObamaForPrison & 44\\
50 & QAnon & 44\\
51 & Censorship & 43\\
52 & CrimesByRace & 40\\
53 & Jewspiracy & 40\\
54 & russia & 40\\
55 & War & 39\\
56 & Kikes & 38\\
57 & Abortion & 37\\
58 & MassDeport & 37\\
59 & Israel & 36\\
60 & presstitutes & 36\\

\end{longtable}

\clearpage

\begin{table*}[htbp]
\centering
\caption{\textbf{Percentage of domains matched with MBFC ratings for each platform.}}
\begin{tabular}{lr}
\label{tab:percentage_rated}
Platform & \% of rated domains\\
\midrule
Facebook & 5.85\\
Reddit & 26.62\\
Twitter & 3.94\\
YouTube & 11.11\\
\addlinespace
Bitchute & 18.84\\
Gab & 27.77\\
Parler & 18.98\\
Scored & 22.06\\
Voat & 34.12\\
\bottomrule
\end{tabular}
\end{table*}

\clearpage

\begin{figure*}[h]
  \centering
  \includegraphics[width=\textwidth]{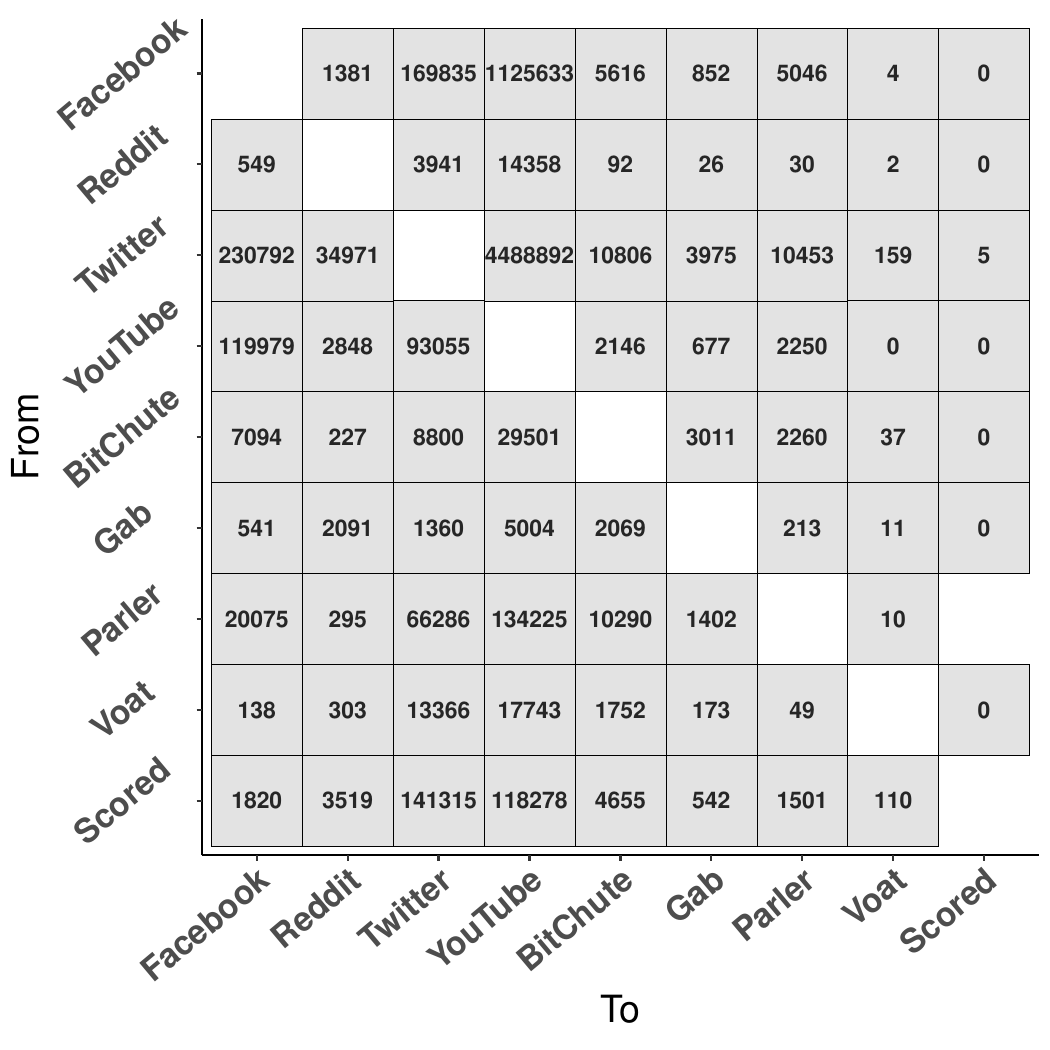} 
  \caption{\textbf{Adjacency matrix describing the number of links from and to each of the nine different platforms.}}
  \label{fig:adjacency_matrix}
\end{figure*}


\begin{figure*}[h]
  \centering
  \includegraphics[width=\textwidth]{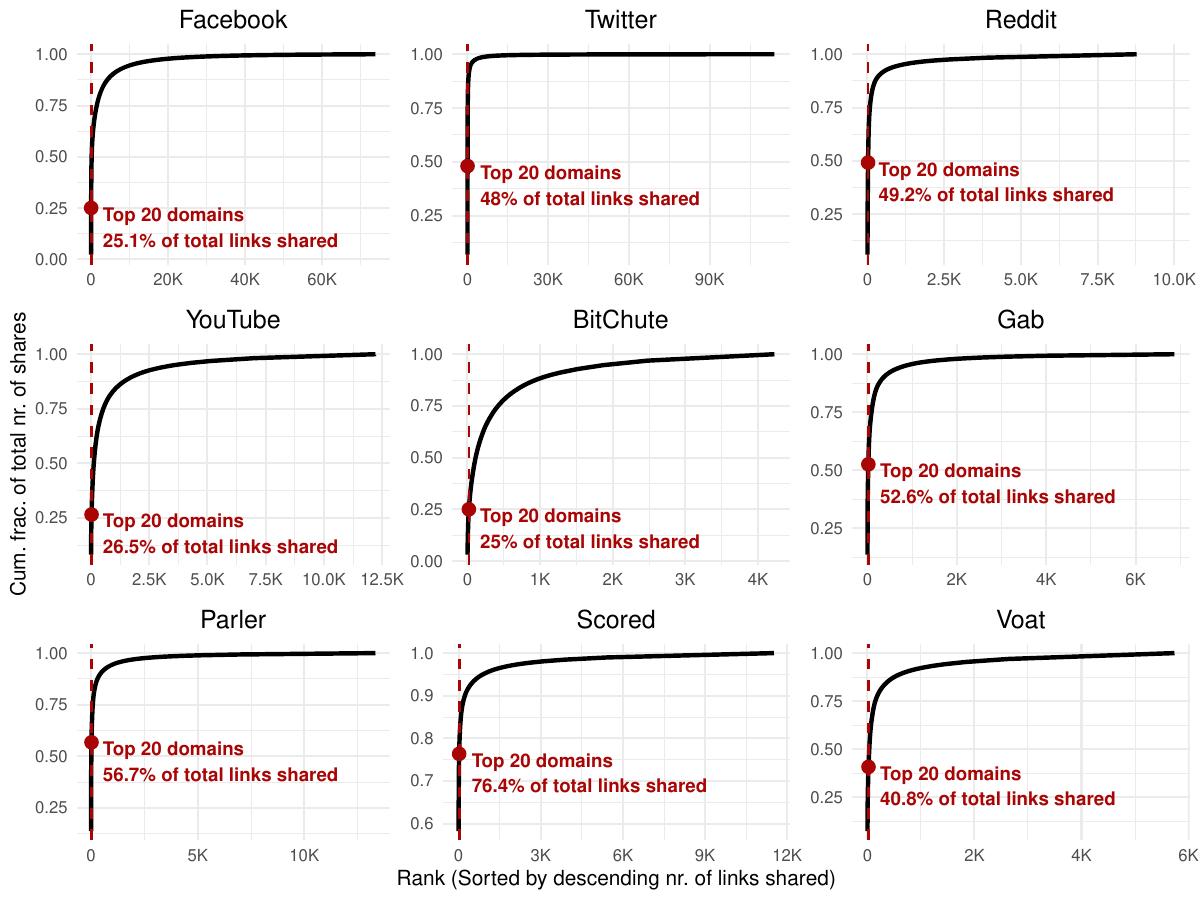} 
  \caption{\textbf{Cumulative fraction of the total amount of links towards different domains. As observed, the top-20 most shared domains encompass for all platforms a significant percentage of links shared.}}
  \label{fig:top20}
\end{figure*}


\begin{figure*}[h]
  \centering
  \includegraphics[width=\textwidth]{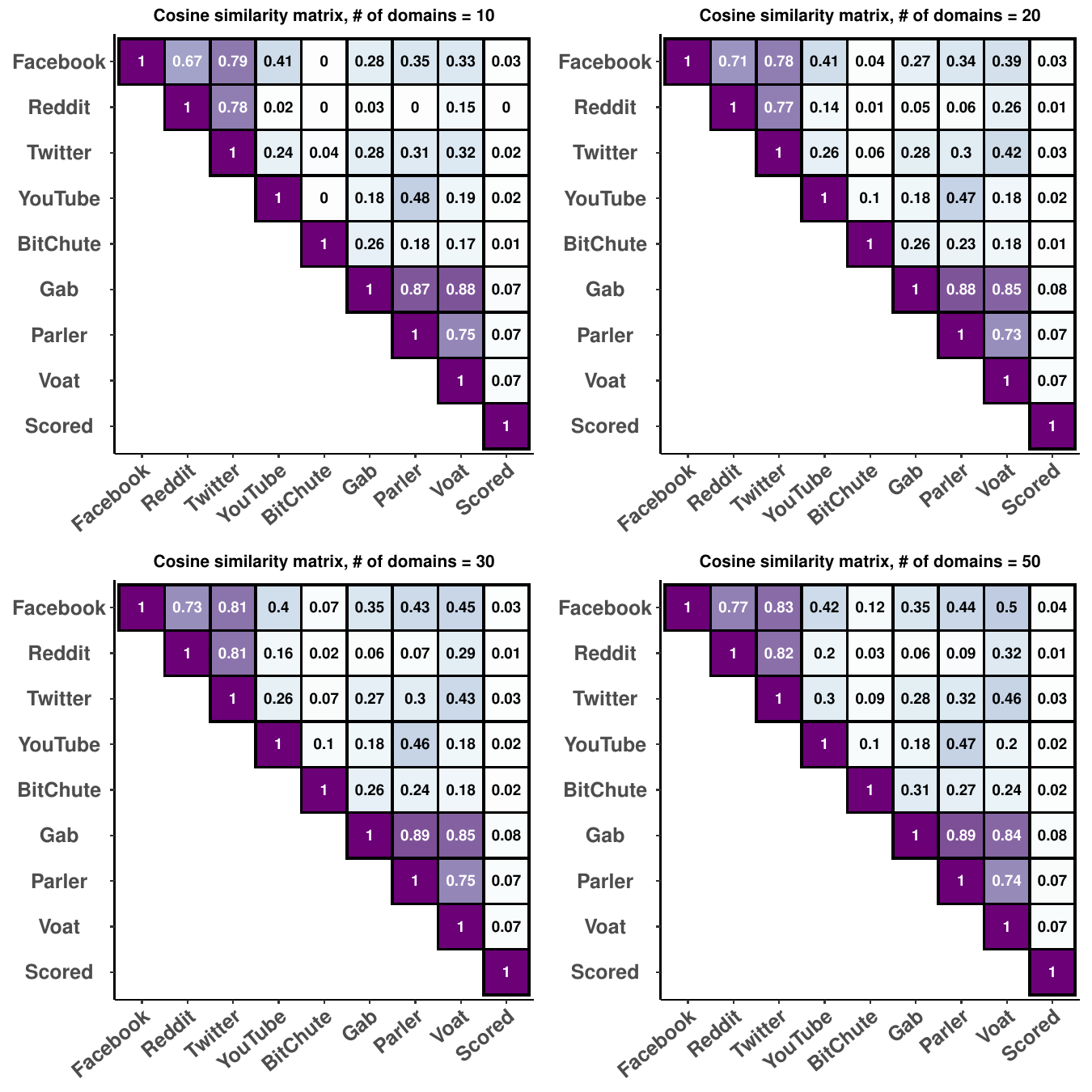} 
  \caption{\textbf{Cosine similarity matrix for different thresholds (top 10, 20, 30, 50 domains). The results remain extremely similar between the different matrices.}}
  \label{fig:cosine_matrix}
\end{figure*}


\begin{figure*}[h]
  \centering
  \includegraphics[width=\textwidth]{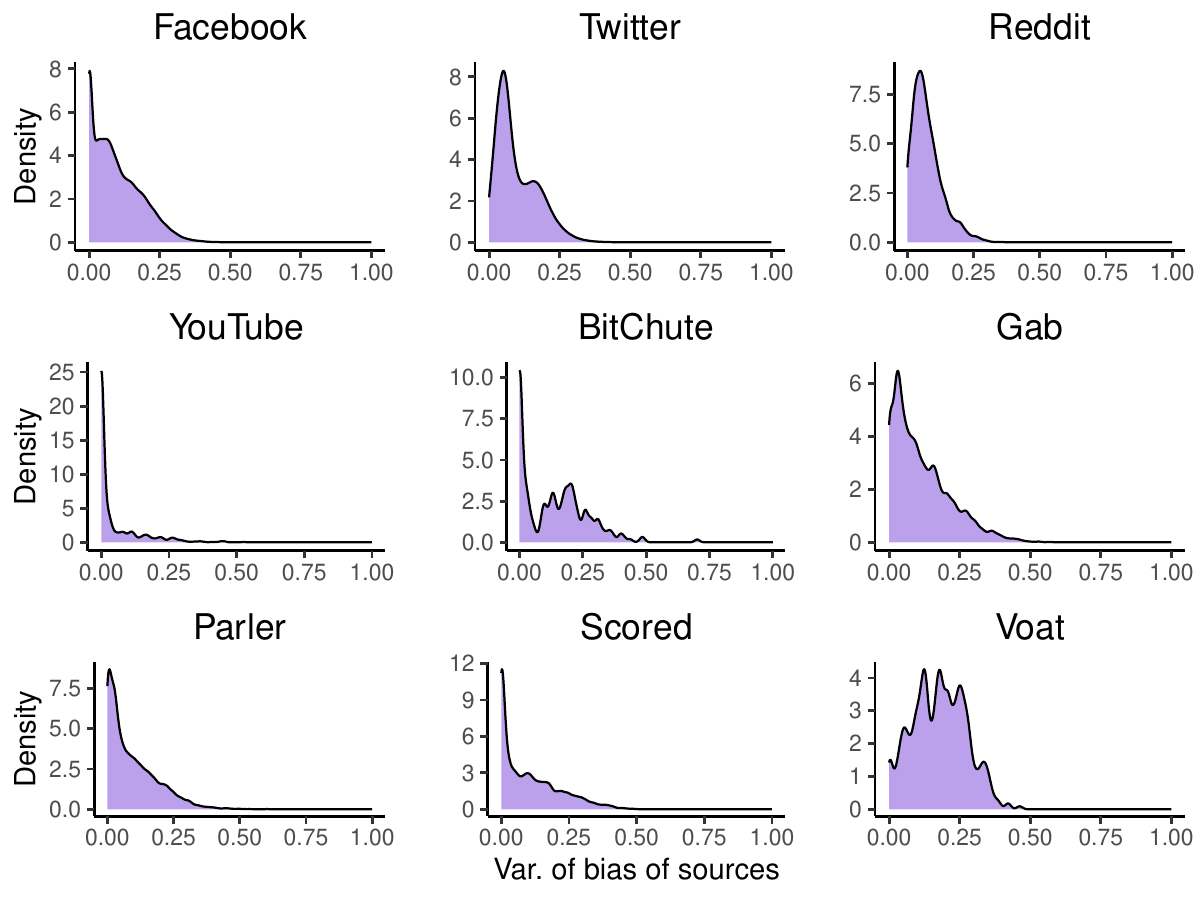} 
  \caption{\textbf{Probability density function of the variance in the bias of the sources shared by users on different platforms. We observe that users on different platforms generally exhibit low variance in the political bias of the sources they share, suggesting that most tend to engage with content from a narrow ideological range.}}
  \label{fig:cosine_matrix}
\end{figure*}


\begin{figure*}[h]
  \centering
  \includegraphics[width=\textwidth]{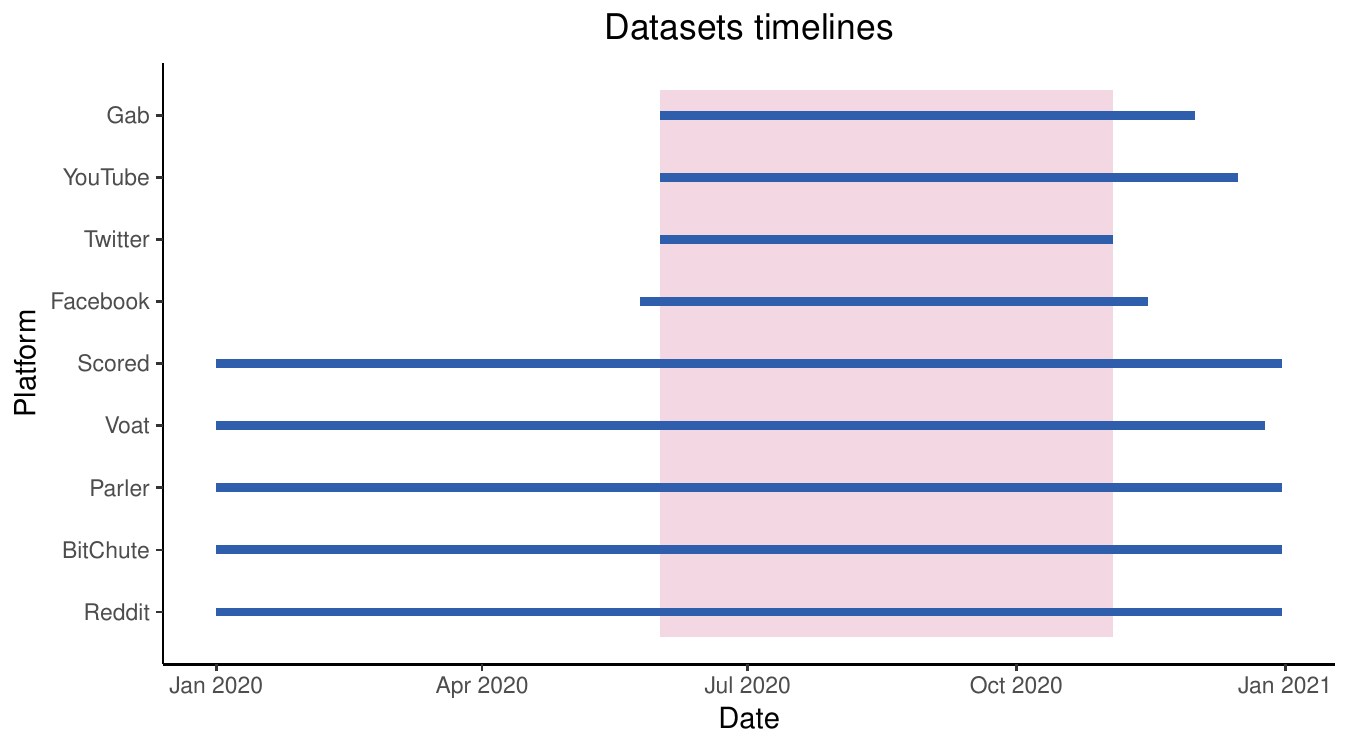} 
  \caption{\textbf{Common time window of the data sets.}}
  \label{fig:cosine_matrix}
\end{figure*}

\begin{figure*}[h]
  \centering
  \includegraphics[width=\textwidth]{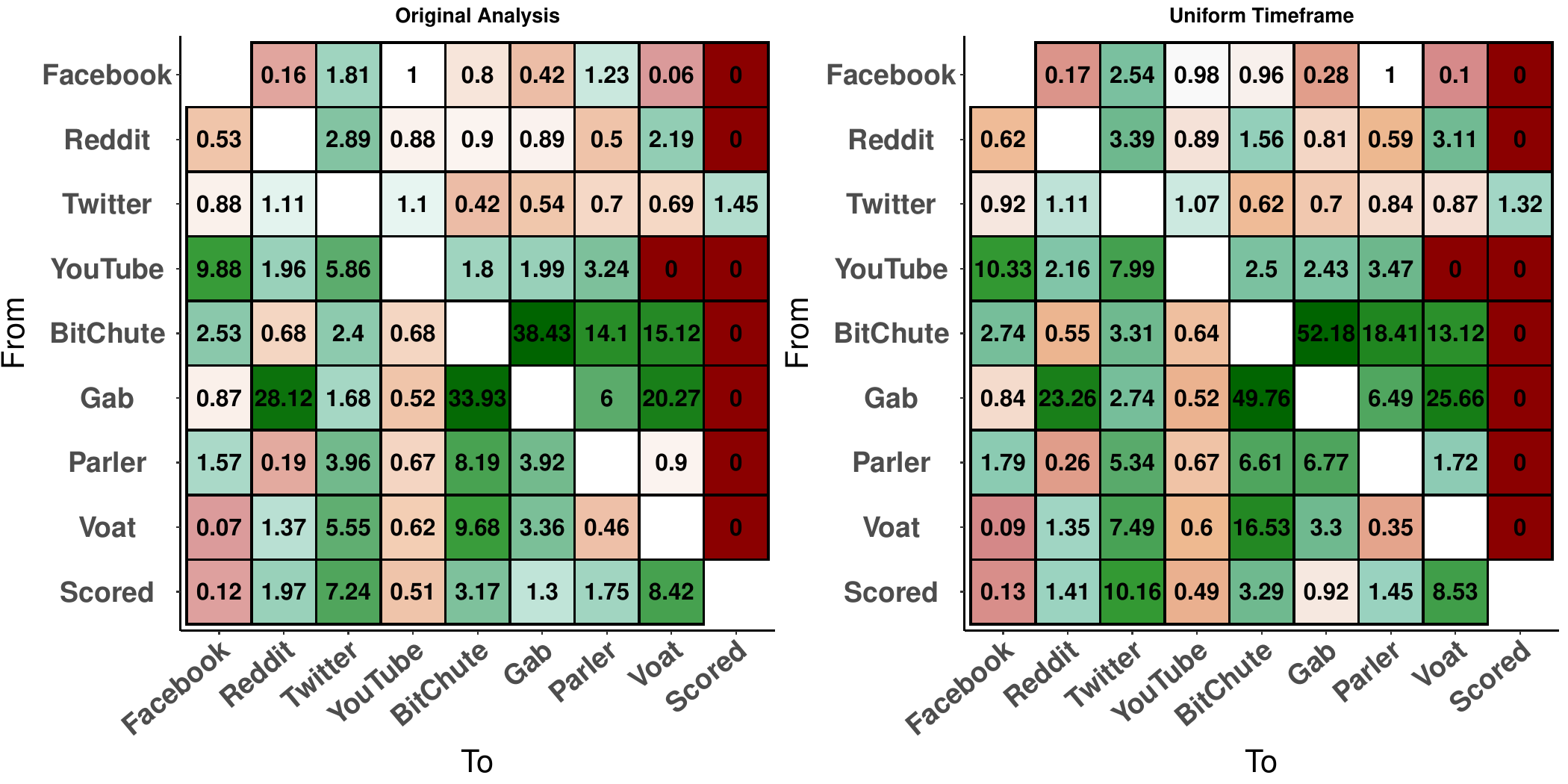} 
  \caption{\textbf{comparison between the rescaled adjacency matrix computed with the “full” datasets, and those obtained after filtering the datasets on the common time window. The results remain largely consistent between the two, both highlighting a fringe ecosystem generally more connected than the mainstream one. This similarity between the results is confirmed by computing Kendall’s Tau between the two matrices, which has a very high score of $\tau = 0.92$.}}
  \label{fig:cosine_matrix}
\end{figure*}


\begin{figure*}[h]
  \centering
  \includegraphics[width=\textwidth]{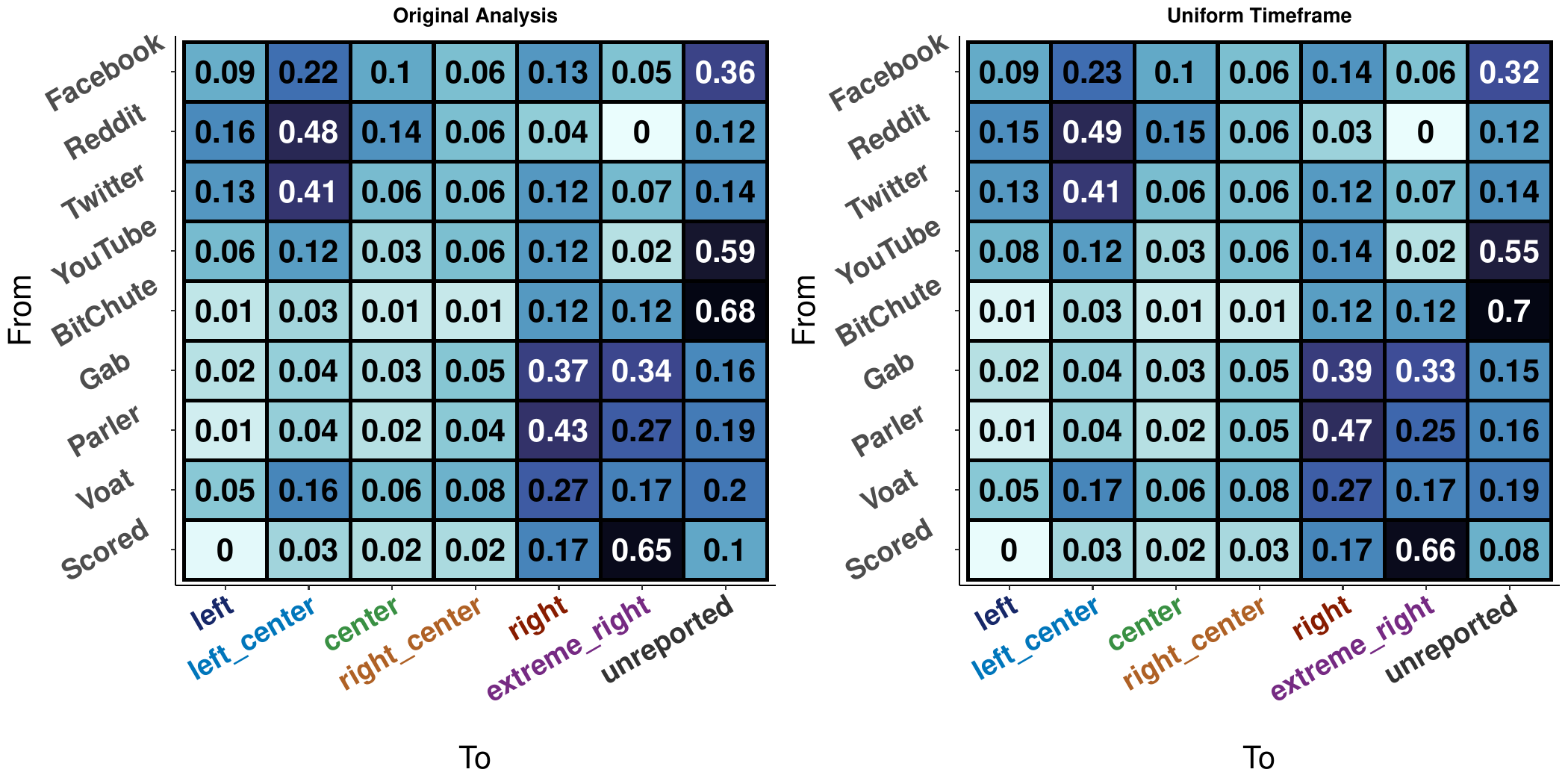} 
  \caption{\textbf{Comparison of the original platform diet to the one computed after filtering the data sets maintaining the uniform time window. The results are practically the same, as also shown by the extremely high Kendall’s Tau value between the matrices of $\tau = 0.967$.}}
  \label{fig:cosine_matrix}
\end{figure*}


\begin{figure*}[h]
  \centering
  \includegraphics[width=\textwidth]{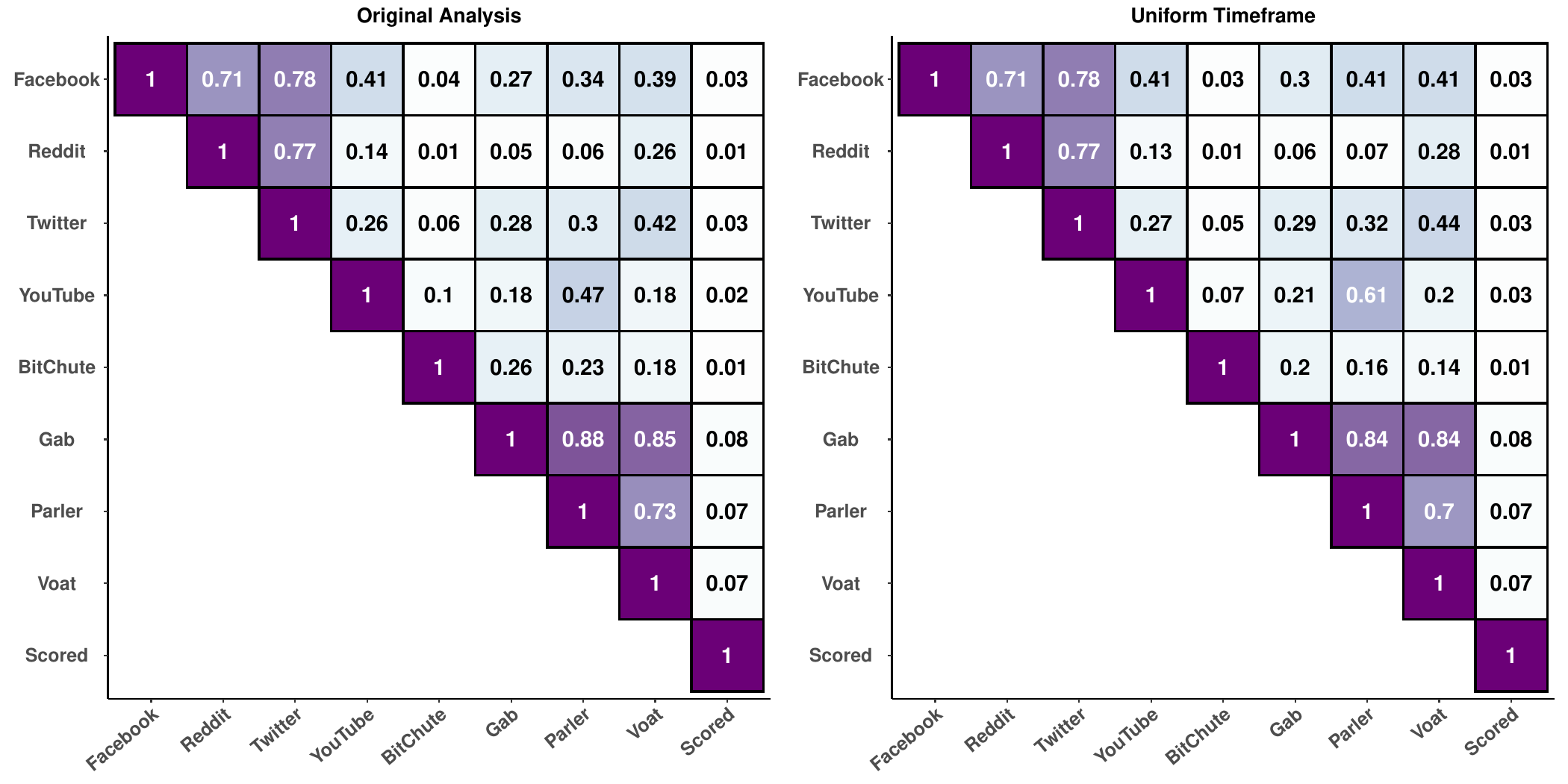} 
  \caption{\textbf{Cosine similarity matrix computed on both the original and filtered dataframes. The values between the two are extremely similar, as also shown by the high value of Kendall’s Tau between the matrices ($\tau = 0.966$). The only notable difference is an increase in the similarity between YouTube and Parler.
}}
  \label{fig:cosine_matrix}
\end{figure*}


\begin{figure*}[h]
  \centering
  \includegraphics[width=\textwidth]{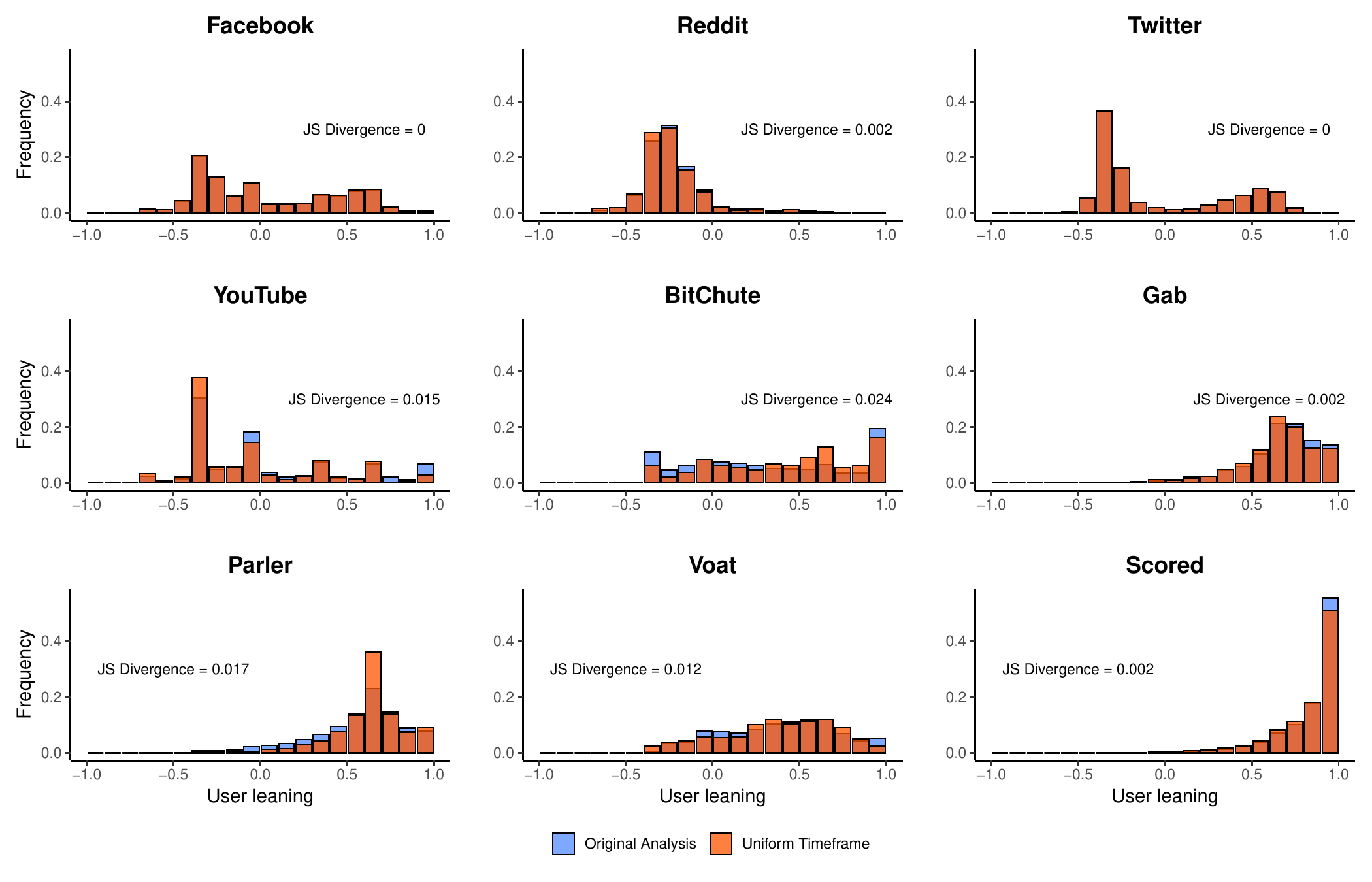} 
  \caption{\textbf{Overlap between original user leaning, and those obtained after filtering the data. We can observe that the histograms nearly completely overlap. This similarity is also shown by the reported values of the Jensen-Shannon divergence between the histograms, which results very low in every case considered.}}
  \label{fig:cosine_matrix}
\end{figure*}

\clearpage


\end{document}